# Waiving Article Processing Charges for Least Developed Countries. A Brick Stone of a Large-scale Open Access Transformation

Niels Taubert (https://orcid.org/0000-0002-2357-2648)
Andre Bruns (https://orcid.org/0000-0002-2976-0826)
Christopher Lenke (https://orcid.org/0000-0002-9232-4736)
Graham Stone (https://orcid.org/0000-0002-5189-373X)

## Abstract

This article investigates the question, if it is economically feasible for a large publishing house to waive article processing charges for the group of 47 so called least developed countries (LDC). As an example Springer-Nature is selected. The analysis is based on the Web of Science, OpenAPC and the Jisc collections Springer compact journal list. As a result, it estimates an average yearly publication output of 520 publications (or a share of 0.26% of the worldwide publication output in Springer-Nature journals) for the LDC country group. The loss of revenues for Springer-Nature would be 1,1 million $ if a waiver would be applied for all of these countries. Given that money is indispensable for development in the case of LDC (e.g. life expectancy, health, education), it is not only desirable but also possible in economic terms for a publisher like Springer-Nature to waive APCs for these countries without much loss in revenues.

## I. Introduction

In recent years a number of funding organizations and research councils have started to support a large-scale transformation towards gold open access (OA) that is based on article processing charges (APC). National-wide OA-contracts were negotiated for Austria, Finland, Hungary, Germany, the Netherlands, Norway, Poland, Qatar, Sweden and the UK.[1] The transformation of the publication environment from a subscription-based towards APC-based financial model is undoubtedly complex and bears some risks for all parties involved. In common with a publication market based on subscriptions; the APC-model comes with some challenges that protagonists must respond to. A major problem of the subscription model has been the limited access to scientific information, especially at locations where funds are scarce and perpetually increasing prices result in a library or serials crisis. In an APC-based environment, different challenges such as questionable publication practices (Bohannon 2013), double-dipping (Prosser 2015, Pinfield et al. 2017) and a redistribution of financial burdens of the publication system (e.g. Smith et al 2016) arose and were solved at least in part.

One aspect that recurrently comes up in the political discussion about open access is the question on how to deal with APC in the case of countries in the so-called Global South. It is likely that many do not have enough public funds for research to cover the costs for APC or have other priorities than establishing structures and workflows for the organization of payments for APC. If a large-scale transformation towards APC-based OA would occur on a global level, the risk is that the patterns of exclusion might change. In a subscription-based publication environment, readers in countries of the

---

[1] https://www.springer.com/gp/open-access/springer-open-choice/springer-compact, (accessed on July 2nd 2020).

Global South tend to be excluded from access to published research because of paywalls and lack of public funds for subscriptions. In a gold open access environment based on APC, authors from countries of the Global South might be excluded because of lack of funds for publishing. Some publishers have already responded to that challenge by waiving APC in some of their journals for reprint (RP) authors coming from such countries.[2] Given that waivers are usually applied to full OA journals and given that hybrid OA journals are excluded, current models do not provide a comprehensive solution from countries of the Global South.

This article analyzes the possibility of a waiver of APC for countries of the Global South from an empirical perspective, focusing on one of the large publishing houses: Springer-Nature. It estimates how many publications would be affected if Springer-Nature decided to waive APC in all of their journals as well as the loss of revenues that would result from such a step.

The identification of countries as 'poor' and notions like 'Global South' bear normative implications and the act of attributing such classifications may be contested, undesired and may ill reflect the self-image of these states. An analysis like this can hardly escape this problem as it necessarily must draw on some kind of classification to identify countries where a waiver of APC would be reasonable. For the purpose of this study the country classification of 'least developed countries' (LDC) seems to be suitable. LDC is a country classification applied by the Committee for Development Policy (CDP) of the United Nations (UN). Unlike the World Bank classification of countries into low, lower-middle, upper-middle and high income countries, that is an obvious alternative, the LDC classification is not based on one (economic) criterion only but on a combination of three: Income, human assets index, and economic vulnerability index. A recommendation for inclusion takes place if a country does not meet a certain threshold in one of the three criteria, a graduation takes place if a country falls below a higher threshold of two of the three criteria. Income is defined as gross national income per capita and an inclusion in the LDC requires a three-year average lower than $1,025.[3] The Human Assets Index (HAI) is a composite index including the health indicators 'under-five mortality rate', 'percentage of population undernourished', 'maternal mortality ratio' and the two education indicators 'gross secondary school enrolment ratio' and 'adult literacy rate'. The economic vulnerability index is also a composite index that intends to measure structural vulnerability to economic and environmental shocks and is composed of eight indicators.[4] In 2013 0.7% of global researchers were located in LDC and were involved in 0.6% of the worldwide publication output (UNESCO 2015). The most recent LDC list with 47 countries published in 2018 is used for this study.[5]

## II.     Literature Review

Besides its relevance in a political and bargaining context, this article contributes to a growing field of studies that aim to analyze the current transformation process towards Gold OA publishing based on APC (Solomon and Björk 2012, Björk and Solomon 2015). Their goal is to understand both the dynamics of the market and the economics of the publishing model.

---

2   Wiley, for example, applies a pricing model with waivers and discounts for some countries and some journals https://authorservices.wiley.com/open-research/open-access/for-authors/waivers-and-discounts.html, Springer-Nature and BioMedCentral waives APC for low income countries and offers discounts for lower middle income countries https://authorservices.wiley.com/open-research/open-access/for-authors/waivers-and-discounts.html, https://www.springernature.com/de/open-research/policies/journal-policies/apc-waiver-countries but only for full OA and not for hybrid journals (accessed on July 2nd 2020).

3   This is also the threshold of the World Bank for including a country in the group of LIC.

4   The LDC Identification criteria and indicators are published on the website of the UN: https://www.un.org/development/desa/dpad/least-developed-country-category/ldc-criteria.html (accessed on July 2nd 2020).

5   The list of countries used in this report was retrieved from the website of the UN https://www.un.org/development/desa/dpad/least-developed-country-category/ldc-data-retrieval.html (accessed on July 2nd 2020).

By no means all journals providing immediate OA charge an APC. Journals that do not charge publication fees are sometimes called platinum (Wilson 2007) or diamond OA (Fuchs and Sandoval 2013). Firstly, at the global level, Morrison et al. (2015) find that more than two thirds of the journals included in the Directory of Open Access Journals (DOAJ)[6] apply publication fees. The application of APC seems to differ by field (Crawford 2017). For example, for medicine two thirds of the journals refrain to impose APC (Asai 2019). In addition, the take-up of APC also varies by region. A large share of OA journals not charging APC can be found in Latin America, the Middle East, and Eastern Europe (Crawford 2017). They are financed by other means, such as subsidies from the state as in the case of Brazil (Appel and Albagli 2019) or they are driven by the voluntary and unpaid work of dedicated scientists.

Second, a number of studies are interested in the dynamics of the transformation to OA and address to what extent the publication output of an entity of a research system (e.g. institutions, countries, disciplines) is freely available online via the formal communication channel.[7] Studies differ with regard to the databases and the sources of OA information being used as well as the definition of OA types, (Martín-Martín et al. 2018) but, nevertheless, there is some evidence that can be found across all contributions: The share of publications that are freely available online in the formal communication channel has reached a level that can hardly be overlooked and that today contributes to the supply of information within many fields of the sciences, the social sciences and the humanities. In addition, the dynamics of growth of the Gold OA share still sustains.

A third set of studies is interested in the price for publishing in an APC environment. Because of the lack of other data, early studies referred to list prices on publishers' websites (Morrison et al. 2015) or to prices as recorded by DOAJ (Björk and Solomon 2015). Given that the amount of money that is actually paid for APC can differ from list prices, and given that payments for articles published in the same journal may also vary, more recent studies are based on collections of actual payments (Jahn and Tullney 2016). Five results of the APC prices/payment studies seem to be worth highlighting: The average price/payment reported different studies varies at a similar scale between €905 (Asai 2019) and €1,479 (Pieper and Broschinski 2018). All studies report large standard deviation indicating that there is much variance in the pricing of APC. In addition, there is some evidence that APCs are higher for hybrid journals than for full open access journals[8] and that APC varies by discipline, (Solomon and Björk 2016) type of publisher (Asai 2019), quality (Björk and Solomon 2015) and language of the journal (Asai 2019).

A fourth tier of studies investigate the financial effects of an ongoing OA transformation on the level of institutions showing that the transformation towards an APC model might overburden the library budget of research-intensive universities (Solomon and Björk 2016, Taubert 2019).

## III. Research Question

The aim of this analysis is to investigate the financial consequences of waiving APC for authors from the Global South. Without any doubt, such a move would help the publisher`s reputation within the scientific community and might be an option worth considering. However, costs in terms of loss of revenue must be clear to the publisher from the outset.

---

[6] https://doaj.org/ (accessed on July 2nd 2020).

[7] Laakso et al. 2011, Gargouri et al. 2012, Archambault et al. 2014, Crawford 2015, Wohlgemuth et al. 2017, Piwowar et al. 2018, Abediyarandi and Mayr 2019, Hobert et al. 2020.

[8] Pinfield et al. 2016, Jahn and Tullney 2016, Schönfelder 2020.

This article answers this question for Springer-Nature journals covered by the UK Springer Compact Agreement. Springer-Nature was chosen as it is one of the largest publishing houses worldwide with a strong engagement in OA publishing. The UK Agreement was selected as a case as it collects the majority of Springer's journals that apply a hybrid open access model and are of strategic importance for a transformation towards APC-based OA. The Springer Compact Agreement 2016–2018 includes 1,997 Springer-Nature journals, covering all fields in the sciences, social sciences and humanities and allows all members of participating institutions to publish their articles OA.

The empirical analysis is organized in two steps. In order to determine the volume of revenues for Springer-Nature in a possible future APC-based publication market, the distribution of reprint (RP) authors, sometimes also called 'corresponding authors', in all journals covered by the Springer Compact Agreement is analyzed by country. Identifying the RP author of a publication is important in economic terms as it is assumed that the RP authors' institution should pay for the publication in an APC-based publication market. After an overview of the worldwide distributions, the numbers of publications with RP authors (in what follows: 'RP publications') is calculated for the LDC country group, as well as for each individual country. In addition, estimations of potential losses of revenues are reported.

## IV. Methods

### Data Sources

The analysis makes use of three data sources:

- *Jisc Collections Springer Compact 2016-2018*: a list containing Springer-Nature journals was used to identify the relevant set of publications for this study.
- *Publication database*: publication data and reprint author information were taken from the Web of Science (WoS). Raw data from WoS were provided by the Competence Centre for Bibliometrics.[9] The processed raw database in its version of February 26th 2020 was used in order to conduct an up-to-date analysis. WoS data allows the numbers of RP publications to be determined for each country in the list of Springer Compact journals as far as they are covered by WoS.
- *APC cost information*: In order to obtain the costs for APCs that were actually paid by institutions the OpenAPC dataset was used. It is the largest collection of APC payment information from various countries.[10] OpenAPC was also used for an estimation of the number of publications not covered by WoS and the calculation of a correction factor.

### Data analysis

As a first step, a table with all article-address-combination was created for all citable items in journals of the Jisc Collections Springer Compact 2016–2018 list. 'Citable items' include the publication types 'article', 'review', and 'proceedings paper'[11] for which APC are usually paid (Bruns et al. 2019). The time span covers publications from 2016 to 2018. Electronic and print ISSN was used for matching the Springer Compact list with WoS.

The second step was to calculate the number of RP publications for each country (the table was enriched with additional country information). In cases in which a publication had more than one reprint author from different countries, the publication was counted for each country so as not to underestimate the number of publications for a possible waiver. Finally, loss of revenue for Springer-Nature as a consequence of waiving APC were calculated for LDC as well as for each country in the categories.

---

[9] http://www.forschungsinfo.de/Bibliometrie/en/index.php?id=home (accessed on July 2nd 2020).

[10] On March 31st 2020 it contained cost information 104,661 OA articles in full OA and hybrid journals, amounting to € 207,687,858 and contributed by 262 institutions (https://www.intact-project.org/openapc/, accessed on July 2nd 2020).

[11] http://help.incites.clarivate.com/incitesLiveJCR/9607-TRS (accessed on July 2nd 2020).

## V. Results

### Overview

What would a publication market based on APC look like and from which country would the bulk of revenues for Springer-Nature come from? An overview of the worldwide distribution of RP publications by country is given below. Graph 1 shows a scatter plot of the distribution of all countries worldwide ordered by the gross national income in million $ and the number of publications in the period 2016–2018 with a reprint author from that country. The two countries with the largest publication output in Springer-Nature journals, China and the United States with 88,278 and 65,376 RP publications respectively, were excluded for better visualization. The distribution already indicates that there are a relatively small number of countries with a strong RP publication record where the lion's share of Springers-Nature's income would come from. The group of least developed countries can hardly be detected in the lower left corner as their gross national income and their number of RP publications in 2018 are both small.

*Figure 1: Countries with < 50,000 RP publications between 2016–2018, by GNI*

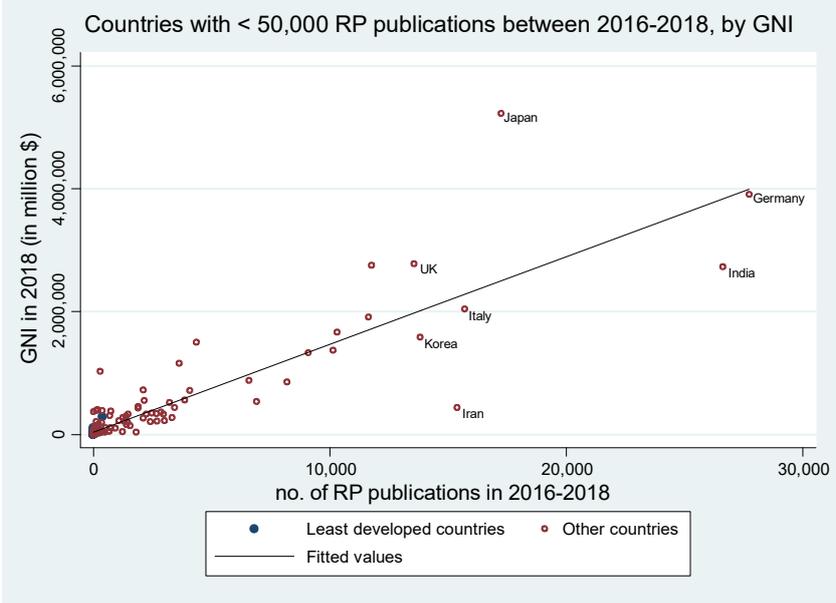

Graph 2 zooms in and plots countries with an RP publication output of less than 500, which makes the group of the least developed countries visible. With the exception of Bangladesh and Ethiopia the number of RP publications is smaller than 100, thus indicating that this group does not currently contribute much to an APC-based publication market. In addition, it is interesting to note that a considerable number of these countries have an RP publication output of less than five publications.

*Figure 2: Countries with < 500 RP publications between 2016–2018, by GNI*

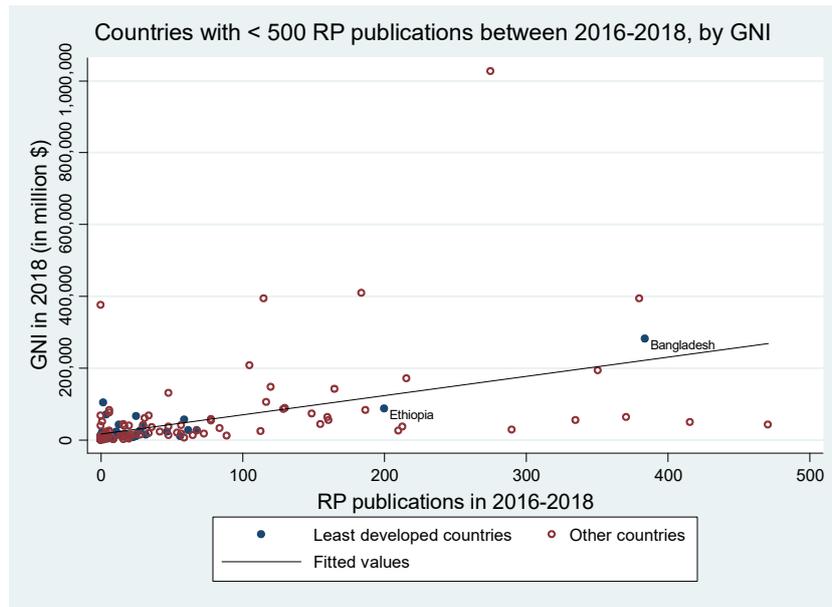

Graph 3 orders countries again by number of RP publications in the period 2016–2018 but now by GNI by capita. Two results seem to be worth noting: First, there are countries with a strong RP publication record but with a relatively low GNI per capita. The most prominent example is India with a GNI per capita of $2,020 in 2018. Second, there are some countries with a very high GNI per capita with little or no publication output. Examples are Macao, Luxembourg, Hong Kong or The Bahamas.

*Figure 3: Countries with < 50,000 RP publications between 2016–2018, by GNI per capita*

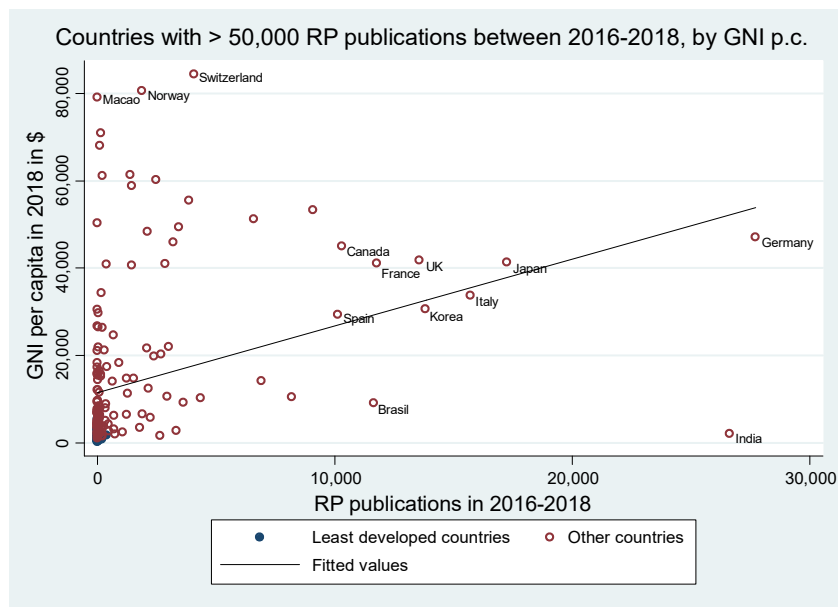

Again, the group of LDC can be hardly detected in the graph. The zoom (graph 4) shows that the GNI per capita is far below $5,000 for most of the LDC group with the exception of Tuvalu. The two

countries with the strongest publication output in the LDC group both have a low GNI per capita ($1,750 in the case of Bangladesh and $750 in the case of Ethiopia).

Figure 4: *Countries with < 500 RP publications between 2016–2018, by GNI per capita*

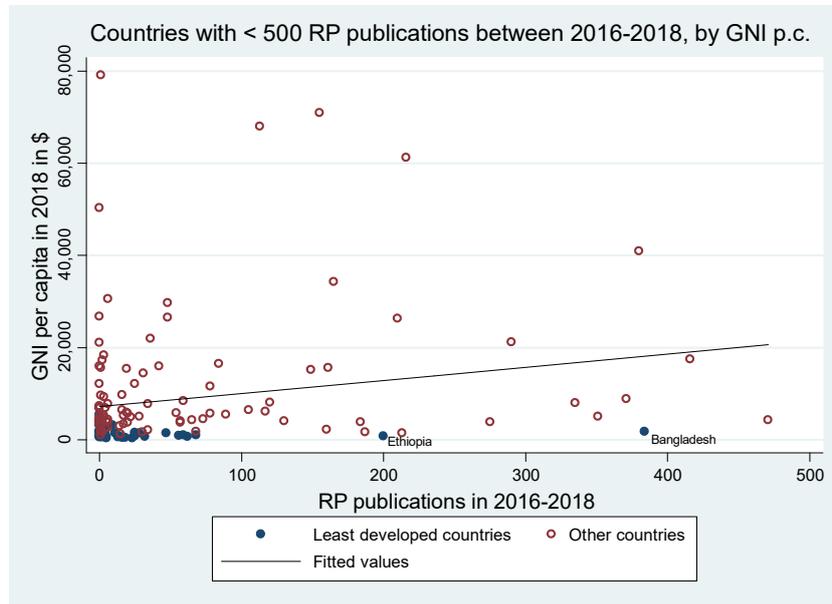

## Correction Factors

The analysis of the number of RP publications by country based on WoS provides a good overview of the relative share of all countries in a Gold OA publication market, when APCs are applied. When it comes to the calculation of financial effects of a possible waiver for APC, two shortcomings of the data should be considered: The incomplete coverage of Jisc Collections Springer Compact journals in WoS and the incompleteness of reprint information. In order to overcome both shortcomings and to come to a qualified estimation of possible financial effects, two correction factors are calculated.

*Incomplete coverage of Jisc Collections Springer Compact list in WoS*: WoS covers more than 24,000 journals but is not exhaustive. The matching of the Springer Compact list with WoS revealed that only 1,446 of the 1,997 journals were indexed in WoS. In other words, 551 journals or a share of 28% are not covered. Given that journals differ regarding the number of citable items published, the share of journals is not an adequate correction factor. Therefore, a different approach is undertaken. For UK institutions, the OpenAPC data set comprises all expenditures for APC, including those of the Springer agreement. For this set of publications, the period 2016–2018 was analyzed in order to calculate to what extent they are covered by WoS. The correction factor is simply the ratio of all UK publications in journals of the Jisc Collections Springer Compact list and the number of them covered by WoS.

Table 1: *Number of UK publications in Springer Hybrid Journals in- and outside WoS (Source: OpenAPC, period 2016–2018)*

| *No. of RP publications from UK in Springer Hybrid OA journals* | *No. of them covered by WoS* | *No. of them not covered by WoS* | *Correction factor* |
|---|---|---|---|
| 10,891 | 8,613 | 2,278 | 1.26448 |

*Reprint Information in WoS*: Reprint information of a publication can sometimes be problematic. On the one hand, there are a number of publications with more than one reprint address. In the analysis a pragmatic solution was followed, and all publications were fully counted for all countries involved. This was to counterbalance the number of publications where RP information is missing. In order to consider these publications, the ratio between all publications and those with RP information was calculated as a correction factor.

*Table 2: Number of publications in Springer Hybrid Open Choice Journals covered by WoS: All, with and without RP information (period 2016–2018)*

| No. of publications in Springer Hybrid OA journals | No. of them with RP information | No. of them without RP information | Correction factor |
|---|---|---|---|
| 464,483 | 443,064 | 21,419 | 1.04834 |

# RP publication output of least developed countries

The results of the analysis for the group of least developed countries are given in table 3. The column 'RP pub. all' reports the number of publications of reprint authors from a particular country for the period 2016–2018 in WoS, followed by three columns that break down the number to individual years. Column 'RP pub av.' contains the arithmetic mean of the three years and column 'RP pub. av. corr.' multiplies the arithmetic mean with the two correction factors and can be regarded as a qualified estimation of the overall RP publication output of a country or country group in Jisc Collections Springer Compact journals. The column 'Loss of rev.' calculates the losses of revenues for Springer-Nature in the case that the publisher decides to waive APC for the particular country. It is based on the 'RP pub. av corr.' multiplied by the average amount of APC paid by UK institutions for Springer hybrid journals in 2018. This amount is €2,200 (Marques and Stone 2020).

*Table 3: Least developed countries, number of publications with RP author (2016–2018)*

| Country | ISO3 code | GNI p.c 2018 | RP pub. all | RP pub. 2018 | RP pub. 2017 | RP pub. 2016 | RP pub. av. | RP pub. av.corr. | Loss of rev. |
|---|---|---|---|---|---|---|---|---|---|
| Bangladesh | BGD | 1,750 | 384 | 145 | 138 | 101 | 128.0 | 169.7 | 373,290 |
| Ethiopia | ETH | 790 | 200 | 90 | 69 | 41 | 66.7 | 88.4 | 194,422 |
| Nepal | NPL | 970 | 68 | 20 | 28 | 20 | 22.7 | 30.0 | 66,104 |
| Uganda | UGA | 620 | 62 | 29 | 15 | 18 | 20.7 | 27.4 | 60,271 |
| Tanzania | TZA | 1,020 | 59 | 16 | 24 | 19 | 19.7 | 26.1 | 57,355 |
| Benin | BEN | 870 | 56 | 18 | 21 | 17 | 18.7 | 24.7 | 54,438 |
| Senegal | SEN | 1,410 | 47 | 14 | 12 | 21 | 15.7 | 20.8 | 45,689 |
| Burkina Faso | BFA | 670 | 32 | 13 | 12 | 7 | 10.7 | 14.1 | 31,108 |
| Yemen, Rep. | YEM | NA | 30 | 13 | 8 | 9 | 10.0 | 13.3 | 29,163 |
| Zambia | ZMB | 1,430 | 28 | 10 | 10 | 8 | 9.3 | 12.4 | 27,219 |
| Rwanda | RWA | 780 | 25 | 10 | 5 | 10 | 8.3 | 11.0 | 24,303 |
| Sudan | SDN | 1,560 | 25 | 8 | 9 | 8 | 8.3 | 11.0 | 24,303 |
| Malawi | MWI | 360 | 23 | 8 | 8 | 7 | 7.7 | 10.2 | 22,359 |
| Mozambique | MOZ | 460 | 18 | 8 | 6 | 4 | 6.0 | 8.0 | 17,498 |
| Madagascar | MDG | 510 | 16 | 4 | 8 | 4 | 5.3 | 7.1 | 15,554 |
| Niger | NER | 390 | 16 | 10 | 2 | 4 | 5.3 | 7.1 | 15,554 |
| Congo, Dem. | COD | 490 | 13 | 4 | 3 | 6 | 4.3 | 5.7 | 12,637 |
| Mali | MLI | 840 | 13 | 5 | 4 | 4 | 4.3 | 5.7 | 12,637 |
| Cambodia | KHM | 1,390 | 11 | 4 | 4 | 3 | 3.7 | 4.9 | 10,693 |
| Vanuatu | VUT | 3,130 | 9 | 2 | 4 | 3 | 3.0 | 4.0 | 8,749 |
| Togo | TGO | 660 | 5 | 3 | 2 | 0 | 1.7 | 2.2 | 4,861 |
| Burundi | BDI | 280 | 5 | 5 | 0 | 0 | 1.7 | 2.2 | 4,861 |
| Lao PDR | LAO | 2,450 | 4 | 0 | 2 | 2 | 1.3 | 1.8 | 3,888 |
| Myanmar | MMR | 1,310 | 4 | 3 | 0 | 1 | 1.3 | 1.8 | 3,888 |
| Guinea | GIN | 850 | 3 | 1 | 2 | 0 | 1.0 | 1.3 | 2,916 |
| Bhutan | BTN | 2,970 | 2 | 1 | 0 | 1 | 0.7 | 0.9 | 1,944 |
| Angola | AGO | 3,370 | 2 | 1 | 0 | 1 | 0.7 | 0.9 | 1,944 |
| Mauritania | MRT | 1,160 | 2 | 0 | 0 | 2 | 0.7 | 0.9 | 1,944 |
| Lesotho | LSO | 1,390 | 2 | 1 | 0 | 1 | 0.7 | 0.9 | 1,944 |
| Gambia, The | GMB | 710 | 2 | 1 | 1 | 0 | 0.7 | 0.9 | 1,944 |
| Guinea-Bissau | GNB | 750 | 2 | 0 | 2 | 0 | 0.7 | 0.9 | 1,944 |
| Sierra Leone | SLE | 490 | 2 | 0 | 1 | 1 | 0.7 | 0.9 | 1,944 |
| Solomon Isl. | SLB | 2,020 | 1 | 0 | 0 | 1 | 0.3 | 0.4 | 972 |
| Eritrea | ERI | NA | 1 | 1 | 0 | 0 | 0.3 | 0.4 | 972 |
| Afghanistan | AFG | 550 | 1 | 0 | 1 | 0 | 0.3 | 0.4 | 972 |
| Somalia | SOM | NA | 1 | 1 | 0 | 0 | 0.3 | 0.4 | 972 |
| Liberia | LBR | 610 | 1 | 1 | 0 | 0 | 0.3 | 0.4 | 972 |
| Chad | TCD | 670 | 1 | 0 | 0 | 1 | 0.3 | 0.4 | 972 |
| Djibouti | DJI | 3,190 | 0 | 0 | 0 | 0 | 0.0 | 0.0 | 0 |
| Tuvalu | TUV | 5,430 | 0 | 0 | 0 | 0 | 0.0 | 0.0 | 0 |
| Cent. Afric. Rep. | CAF | 490 | 0 | 0 | 0 | 0 | 0.0 | 0.0 | 0 |
| Timor-Leste | TLS | 1,820 | 0 | 0 | 0 | 0 | 0.0 | 0.0 | 0 |
| South Sudan | SSD | NA | 0 | 0 | 0 | 0 | 0.0 | 0.0 | 0 |
| Sao Tome | STP | 1,890 | 0 | 0 | 0 | 0 | 0.0 | 0.0 | 0 |
| Kiribati | KIR | 3,140 | 0 | 0 | 0 | 0 | 0.0 | 0.0 | 0 |
| Comoros | COM | 1,380 | 0 | 0 | 0 | 0 | 0.0 | 0.0 | 0 |
| Haiti | HTI | 800 | 0 | 0 | 0 | 0 | 0.0 | 0.0 | 0 |
| **SUM** | | | **1,176** | **450** | **401** | **325** | **392.0** | **519.6** | **1,143,202** |

An important question regarding the publication output is whether there are typical subjects and fields in which reprint authors from the LDC group publish. The WoS provides a subject classification that attributes each journal and all of their publications to one (or more) of 256 subjects.

Figure 5: *LDC publication output, by WoS categories*

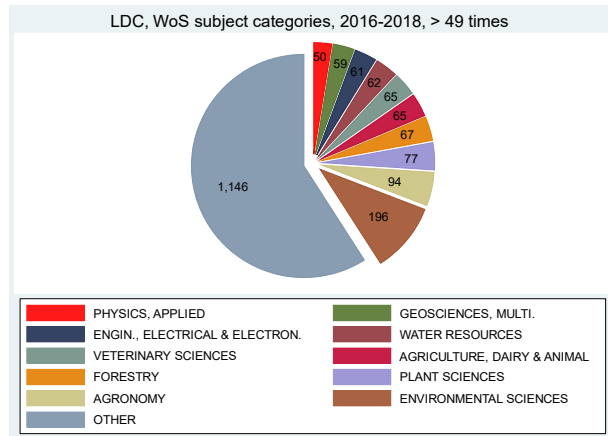

Before the main characteristics of the distribution are described, two methodical remarks should be made. First, RP authors from LDC publish in a large variety of different WoS categories. Therefore, subject categories with an output below a threshold of 50 publications were summarized in the category 'other'. Second, journals can be assigned to more than one of the WoS categories. Therefore, the cumulated number of all subject categories is larger than the number of publications reported in the previous sections.

A look into the distribution reveals that large fractions of the publication output appear to refer to societal conditions, problems and challenges of LDCs. For example, in the subjects 'agriculture', 'agronomy', 'plant science', 'water resources', and 'veterinary science' that are relevant for the production and supply of food, and also for 'environmental science', 'geoscience', and 'forestry' that may study environmental conditions (and changes of these). This distribution indicates that large parts of the research of RP authors from LDC address major societal conditions and provide knowledge of high practical relevance.

## VI. Discussion

This article provides an analysis for the RP publication output of the period 2016–2018 in Springer-Nature journals for the LDC group and for all individual countries within these groups. Given that, on the one hand, the worldwide differences in terms of income are striking and, on the other hand, research and academic publishing are extremely costly activities, the empirical results and the comparisons suggested by this study tend to be absurd. In particular, two empirical results of the study are worth highlighting:

First, it turned out that both the RP publication output in WoS and the estimated overall publication output (in journals in WoS and not in WoS) are low for LDC, when compared with the worldwide publication output.

*Table 4: LDC, GNI and RP publications*

| Country group | Av. GNI p. c.p per country | RP pub. av. | RP pub. av. corr | Share of worldwide RP pub. |
|---|---|---|---|---|
| LDC | $1,345 | 392.0 | 519.6 | 0,26% |

In addition, a skewed publication output is not only be found when comparing countries on a worldwide level but also within the LDC group as Ethiopia's and Bangladesh's share sum up to 50% of the overall publication output of LDC.

Second, the relation of the average costs that are actually paid for a publication in a journal of the Springer Compact list and the average GNI per capita is remarkable. An APC for a single article is much higher than the average income per year of a citizen in an LDC.

Regarding the request for a waiver for APC, the following conclusions can be drawn: The share of RP publications of LDC is low in journals of the Jisc Collections Springer Compact list. It would therefore be possible in economic terms for a publisher like Springer-Nature to waive APCs for these countries without much loss in revenue. Given that money is indispensable for development in the case of LDC (e.g. life expectancy, health, education), it is also desirable that public funds in these countries would not be spent on APC. This particularly applies against the background of the analysis of the subject categories, suggesting that large parts of the publication output are of high societal relevance for LDC. The costs for publications should therefore be covered by other means.

Not all publishers' portfolios are identical and those that specialize in some of the disciplines listed above might see a disproportionate revenue loss. In this case there are various alternative strategies that could be employed based on the particular data set. For example, possible strategies could include the exclusion of certain countries from the waiver, exclusion of certain disciplines, a possible APC discount instead of a full waiver or the number of RP publications, beginning with the country with the largest number. A further model could be for high income countries to cover some of the costs of waivers or reductions. However, it is recognized that some if not all of these scenarios may not be welcomed by the countries in question and this view needs to be balanced against the desire to transition to a fairer open access model.

Whatever the model adopted, waivers and reductions should apply as an automatic procedure and should not require any kind of application by the author. The number of RP publications would need to be monitored to establish a trustful relation between the country and the publisher and to avoid free riding of authors from other countries. OpenAPC is well placed to collate this data on an annual basis and to make it openly available for scrutiny and further analysis. Ultimately, an APC fee waiver for an LDC country would be a temporary solution for as long as a particular country met the conditions outlined above.

## VII. Conclusion

Waiving APCs for LDCs would be a means for publishers to improve their reputation within the scientific community and help them to be attributed as a socially responsible partner of science. In the past, there have been examples of responsible actions of publishers. Besides a waiver for Low Income Countries (LIC) in a set of full OA journals, one may recall the provision of open access to relevant publications in response to the outbreak of swine flu (H1N1) [12] and the current Covid-19 pandemic[13] as well as temporary access to relevant publications in the case of the Ebola crisis in a number of

---

[12] https://open-access.net/en/community/news/article/springer-offers-free-access-to-articles-on-swine-flu, (accessed on July 2nd 2020).

[13] https://www.springernature.com/gp/researchers/campaigns/coronavirus https://www.journals.elsevier.com/journal-of-critical-care/covid-19, https://novel-coronavirus.onlinelibrary.wiley.com/, (accessed on July 2nd 2020).

African countries. [14] The number of publications concerned was at a similar level as the annual publication of LDC countries in Springer-Nature compact journals.

The authors of this article would strongly encourage further empirical research in this area in order to ensure a fair and equitable transition to open access for all countries.

---

14 https://www.elsevier.com/connect/ebola-information-center, (accessed on July 2$^{nd}$ 2020).